\documentclass[11pt]{article}
\usepackage{acl08}
\usepackage{graphicx}
\usepackage{times}
\usepackage{latexsym}
\usepackage{amsmath}
\usepackage[usenames]{color}

\title{A Distributed Method for Trust-Aware Recommendation in Social Networks}

\author{
  Mohsen Jamali \\
  School of Computing Science \\
  Simon Fraser University \\
  {\tt mohsen\_jamali@cs.sfu.ca} %\And
}

\date{}

\begin{document}
\maketitle

\begin{abstract}
This paper contains the details of a distributed trust-aware
recommendation system. Trust-base recommenders have received a lot
of attention recently. The main aim of trust-based recommendation is
to deal the problems in traditional Collaborative Filtering
recommenders. These problems include cold start users, vulnerability
to attacks, etc.. Our proposed method is a distributed approach and
can be easily deployed on social networks or real life networks such
as sensor networks or peer to peer networks.\footnote{This work was
done in 2008 as one of the beginning phases of my PhD thesis.}
\end{abstract}

\section{Introduction}\label{sec:Intro}
With the huge amount of growing information available these days, it
is necessary to have some facilities to help users select the
desired part of information they need. To satisfy this need,
recommender systems have emerged, which mostly used collaborative
filtering approach.

Traditionally, in a recommender system, we have a set of Users $U=
\{u_1, ...\ u_n\}$ and a set of items $I=\{i_1, ...\ i_m\}$. Each
user rates a set of items $RI_u=\{i_{u_1}, ...\ i_{u_k}\}$. The
recommender has the task to recommend some items for the given user
$u$. Generally, Collaborative Filtering approaches for
recommendations try to find users similar to $u$ based on the
similarity of other users' ratings to the ratings of $u$. This is
the primary approach used in recommendations. But users can also
have relations with each other and each user can express a level of
trust to other users (or users he knows). This relations form a
social network called Web of Trust in the literature. This trust can
be used to filter the recommendations based on other users'
preference list. The aim in this project is to investigate how to
inject the concept of trust in recommendations.

The rest of this paper is organized as follows: In section 2 we
discuss a little about the sources of trust and issues in trust
based systems. In section 3, the problem will be formally defined.
We also explain the motivations and challenges faced while dealing
with the problem. Then, related works are discussed in section 4. We
introduce our approach in section 5. The evaluation procedure is
described in section 6. For this project we use a data set which
will be described in section 7. Finally, we present the experimental
results in section 8.

\section{Trust: Where does trust values come from?}

For trust to exist, there must be an expectation in our mind as to a
person's ability to carry out a depended-on action, based upon a
shared set of values . It's unfair to trust that someone will
fulfill your expectations if you don't even know if you share common
values. We trust people as long as they fulfill our expectations.
When they do not, trust can evaporate quickly and take a much longer
time to replace. Where do expectations come from? Expectations come
from values. And where do values come from? We can imagine two types
of sources for trust: explicit expression of trust, and implicit
indication of trust. In some social networks, users explicitly
indicate the users whom they trust
(ePinions\footnote{http://www.epinions.com}). In some of them, users
can even express the level of trust they have on other users
(FilmTrust\footnote{http://trust.mindswap.org/FilmTrust/}).
Recently, the popular social network Facebook has added the
application named "cycle of trust" to its social network in which
people can indicate which users they trust. As We've reviewed the
trust networks, most of them just allow people to indicate whether
they trust other users, or they don't have any trust expressions for
that user. Only a few of them (like FilmTrust) let users express
fuzzy trust values. Another source for trust is implicit trusts
embedded in the social network:
\begin{itemize}
  \item The link structure itself can show trust. When a Webpage has a permanent link to
   another Webpage, it could mean that the author of this page somehow trusts the author of the other page.
  \item In the context of WebPages, the number of page visits for a user $U$'s profile
  from user $V$ can show the trust from $V$ to $U$.
  \item Profile similarity can also show trust. When two users have similar profiles, it means that
  they can trust each other. This way of inferring trust is always subject to the activity of malicious users.
\end{itemize}

\subsection{Trust, Directed or Un-Directed?} From our point of
view, trust is directed. Because you can trust somebody, while
he/she does not trust you. So, this means that trust is different
from friendship which is undirected. You can always trust somebody
while he does not even know you. You trust him just because he is
famous in a topic and his opinions are trusted for you. Notice that
the implicit source of trust, in which we use profile similarity to
infer trust, is an undirected source of trust. Because users with
similar profiles trust each other, and there is no direction.

\subsection{Trust, low Trust, and Distrust} There are different
interpretations for real valued trusts. Mostly, in the literature,
the consider the interval [0,1] for trust. This means that trust
value of 1 is full trust. There are two interpretations for low
trust values. On one hand, some researchers consider low trust
values (values close to zero) as little trust or "don't know"
expression. On the other hand, some researchers consider low values
of trust as distrust. Both approaches have problems. In the first
approach, we can not express distrust on users. In the second one,
we expect distrust values to negatively affect the total trust. But
positive values can not mathematically do that. So, the best choice
is considering the interval [-1,+1] for real trust values. Recently,
there has been a work on propagation of trust and distrust
(\cite{guha_propagation}), which I discussed about it in one the
summaries. Propagation of distrust is an important issue which
should be carefully considered.

\subsection{Issues Affecting Trust}

\begin{center}
\begin{figure}
  \includegraphics[width=7cm]{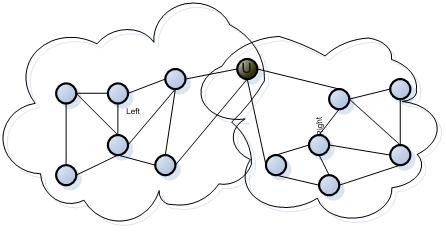}\\
  \caption{Different Trust values for a user in different communities of the network}\label{fig1}
\end{figure}
\end{center}

Generally trust in user $u$ can depend on:
\begin{itemize}
  \item Reference user $v$. Trust in u can depend on the trust of user $v$ (which we have trust to) in $v$.
  \item Community based trust. A social network consists of different
sub-communities. A user can be trusted and well reputed in a
community, while distrusted in another community.  For example as
show in figure \ref{fig1}, user $u$ could be trusted in the left
community, but distrusted in the right community of users.
  \item Another issue which can affect the trust is the topic. A user could be trusted in the network in
   which users rate movies, but the same user could be distrusted in the network in which users rate foods.
\end{itemize}

As the beginning steps of our research we just consider the first
issue. So, we just consider the trust of $u$ to $v$ and transitivity
of trust. We do not take into account the sub-community based or
topic based trusts.

\section{Problem Definition}

Generally, We suppose that recommender systems take some queries as
input, and the recommended items are the output to that query. Four
different (but related) types of queries can be identified in a
trust-based recommendation system:
\begin{enumerate}
  \item Given a user and an item, predict the recommended rating of
  the user on that item.
  \item For a given user, predict a set of most desired
  (recommended) items. This problem is somehow a general form of the
  first problem, but there could be different approaches for the two
  problems.
  \item The user selects some attribute for the desired items, based
  on the values of those attributes the system recommends some items
  satisfying the attributes as much as possible (i.e. in a movie recommender system, trying to recommend
  action drama movies is an example of such a case).
  \item Recommendation can also be used in email filtering.
In this case, the recommender suggest which emails to read and which
of them to filter. It's similar to spam filtering, but using the
user data.
\end{enumerate}

In this paper, we just consider the first type of queries which is
basic and also well noticed in the literature as we discuss later.
Now, it's time to define the problem. The following paragraphs
contain the formal definition of the problem, input to the system,
and output shown to the user.

Basically, we have a set of users $U= \{u_1, ... u_n\}$ and a set of
items $I=\{i_1, ... i_m\}$. Each user rates a set of items
$RI_u=\{i_{u_1}, ... i_{u_k}\}$. Each rating is a real number in
[0,1]. Each user also has explicit trust expression about users
$TU_u=\{u_{u_1}, … u_{u_t}\}$. The trust values are also in range
[-1,1]. In this scale, 1 means full trust, -1 means full distrust,
and 0 means neutral. These trust information form a social network
so called Web of trust, in which each trust expression corresponds
to a weighted edge in the network.

For a given user $u$ and an unrated item $i$, we seek to find the
recommended (estimated) rating of the user $u$ on this item $i$. The
recommended rating should be estimated based on the information
embedded in user ratings (the user preferences) and trust
information. So, in the social network, we look for trusted users
who already expressed a rating (preference) for the item, and
aggregate these ratings to infer an estimated rating of the item for
user $u$.

\subsection{Motivation} With the huge amount of growing
information available these days, it is necessary to have some
facilities to help users select the desired part of information they
need. To satisfy this need, recommender systems have emerged, which
mostly used collaborative filtering approach. These recommenders
still have some problems. A new user which has no rating can not use
such a recommender system. Also collaborative recommender systems
are subject to attacks by malicious users. So, the concept of trust
has been exploited in recommenders to overcome these problems.
Recently, many social networking services (like
Facebook\footnote{http://www.facebook.com}) and even online
marketing Websites (like eBay\footnote{http://www.ebay.com}) are
using the concept of trust to rate users. Ebay has a global
reputation system which users can exploit it to rate other users.

Now, if we have an automatic trust aware recommendation system,
which exploits both trusts and preferences to infer a set of
recommended items for a user, user can easily access the desired
information he is looking for in the system.

There are also some intellectual challenges in this problem. How the
trust propagates in the network? How can we infer indirect trusts?
This is one of the challenges for this problem. There has been some
works done in this topic. All of them have some problems. In the
following sections we'll discuss about it. Another challenge is how
to combine the trust values and preference values. Also one
important challenge is how to use trust values in model based
recommendation. This is a big question which we will not consider
now, and try to solve it in the course of PhD.

Three questions will be raised when we try to deal with this
problem:
\begin{enumerate}
  \item Which preferences (ratings) do we consider?
        \begin{itemize}
            \item Which users do we consider?
            \item Which paths to those users we consider?
        \end{itemize}
  \item What weight do we assign for the rating of each user?
  \item How do we combine them into a recommendation?
\end{enumerate}

To elaborate the questions, let's look at an example. Suppose we
have the following network as shown in figure \ref{fig2}:

\begin{center}
\begin{figure}
\begin{center}
  \includegraphics[width=6cm]{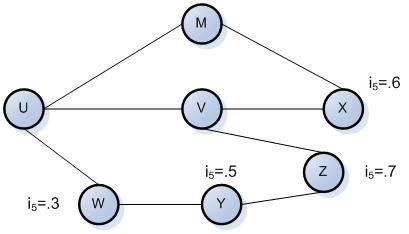}\\
  \caption{A sample Web of Trust with ratings values for an item}\label{fig2}
\end{center}
\end{figure}
\end{center}

In this network, we just show the ratings for item $i_5$. The rating
is shown for the users which have expressed a rating on $i_5$. Now
we want to find out the recommended rating on $i_5$ for user $u$.
Let's review the answers to the three questions above briefly. $u$
has 3 direct neighbor out which only one neighbor ($w$) has
expressed rating. So we consider w into account. Next, $M$ has a
neighbor ($X$) which has expressed rating. So we take X into
account. $V$ also has two neighbors ($X$ and $Z$) which have
expressed ratings. So totally we take $W$,$X$, and $Z$ into account.
Notice that this is a na\"{\i}ve and graphical approach to figure
out which users to consider. We'll discuss the details of our
approach to find the answer later in this proposal.

To answer the second question in this example, we should assign
weights to each the three considered users ($W_x$, $W_w$ , and $W_z$
). In this step, we are not going to cover the solutions for this
question. The general, the idea is to combine the preference
similarity and trust information to infer an impact factor (weight)
for each user.

The 3rd question deals with the combination of the ratings from
selected users into a recommendation. One simple approach is to have
a weighted sum of the ratings from users to get the estimated
rating.

Now, we discuss about the approaches which could be used to answer
the first question. First of all, we should figure out a method to
propagate trust in the network to infer indirect trusts. There has
been some works on this problem, but they all have some issues to be
resolved. So the first thing to do to suggest a solution for
question one is to figure out how trust propagates along the
network. We are not going to the details of what we should do. But
this is an important problem we should resolve as our first step. To
deal with the first question, we can use three approaches:
\begin{figure}
\begin{center}
  \includegraphics[width=7cm]{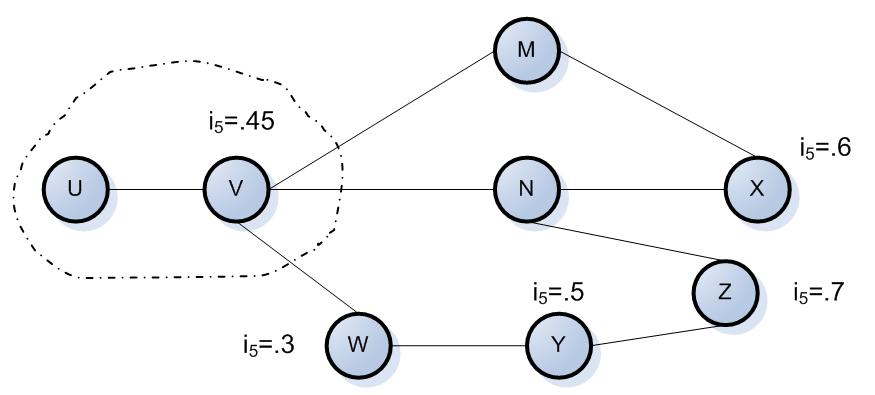}\\
  \caption{In this figure, we just consider user $v$, which is a small set of user to consider.}\label{fig3}
\end{center}
\end{figure}

\begin{enumerate}
  \item \textbf{K-Nearest neighbor approach}

In this approach, we have a fixed constant $k$, which shows the
number of neighbors to be considered. We start from u and walk
through edges between users to visit new users. We select the first
$k$ users which have already expressed ratings for the item we are
looking for the rating. The visit could be done either depth-first
or breadth-first. The depth-first search approach has obvious
problems, since it usually ignores the closer neighbors of $u$. In
the breadth-first approach, also we may loose some neighbors which
are far from $u$ in the network, but the path along $u$ to that
neighbor is almost fully trusted. What we can do is walk through the
network until we reach a user which has expressed rating (This is
the approach used in the example we discussed earlier). This
approach also could lead to some undesired situations as shown in
figures \ref{fig3} and \ref{fig4}.

\begin{figure}
\begin{center}
  \includegraphics[width=7cm]{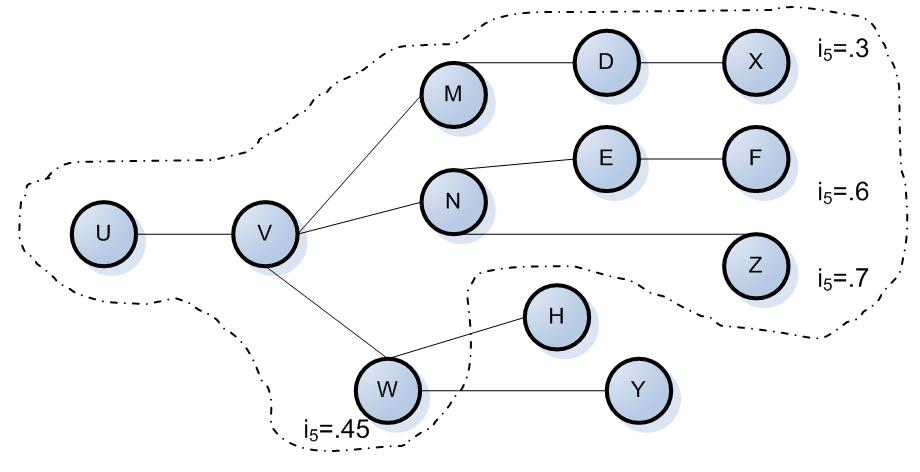}\\
  \caption{In this figure, we have to consider $W$, $X$, $F$, and $Z$. But $X$, $F$, and $Z$ are very far from
   the source user $U$ (this figure is scaled on the length of path from $u$ to for example $X$).
   Notice that the trust along a path is a non-increasing measure,
    so neighbors far from the source are possibly trusted with a low value,
     and to consider them we should walk through long paths which consume time and resource.}\label{fig4}
\end{center}
\end{figure}

  \item \textbf{Trusted users which are trusted above a threshold.}

  What we have in mind is to have a threshold for the inferred trust on the users we visit while we are
  walking through the network. We go deeper in the network until the trust on the user we are visiting drops below
  a certain threshold. Notice that we go deeper regardless of whether the user has express rating or not.

  We should just figure out an appropriate way to infer indirect trust. There are some issues which should be considered:
        \begin{itemize}
            \item How does trust propagate along a path?
            Longer paths should have lower impact. This could be done by a damping factor?.
            Typical trust propagation just use the multiplication of trust values along the path.
            \item How to aggregate the trusts among multiple paths toward a user.
            There has been some works in this area, but they have some fundamental problems.
            One issue we should consider in mind is that if we have for example two
            paths to the target user as shown in figure \ref{fig5}.
            \begin{figure}
            \begin{center}
            \includegraphics[width=5cm]{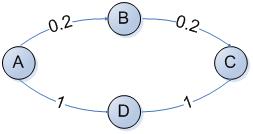}\\
            \caption{In this figure, there are two pathes to target with fundamentally
            different trust values among two paths}\label{fig5}
            \end{center}
            \end{figure}
            In figure \ref{fig5}, one path is a fully trusted path, and the other path is almost not-trusted.
            The fully trusted path tells us that we can trust $C$, but the weakly trusted path will affect
            the total trust (which is not appropriate). So, our approach should be set in a way that
            these paths do not affect the total user very much.
        \end{itemize}

  \item \textbf{Variance based approach}

        An interesting approach could be to consider all
        indirect neighbors of user $u$ which have ratings for the item
        $i$ until the variance among the ratings gets low enough (which means
        the recommendation would be confident enough). There are some issues in this approach. First of all, we can
        not just consider the variance in ratings, because there could be
        some distrusted neighbors which have ratings far from average. We
        should also take the trust into account. So, we should consider the
        variance in a measure which should be an aggregation among trust and
        rating. To have enough neighbors, we could also assign a minimum
        for the size of neighborhood. We should should be really careful while using this approach. When we grow the size
        of neighborhood, it's not necessary that the variance gets lower.
        But, maybe if we consider both trust and rating, since we assume the
        trust has normal distribution, we could get results.

\end{enumerate}

The general question is: how to propagate trust? on one path or
combining multiple paths? Should weakly trusted paths reduce the
trust in a person even if there is a strongly trusted path to that
same person? Considering just one path may cause losing a lot of
information. So, we try to consider multiple paths leading to the
target user. But, the question is how to combine them? As discussed
above, weakly trusted paths should not affect the total trust too
much. One idea for aggregation is using the maximum trusted path.
There also has been a matrix based approach in a paper
\cite{guha_propagation}, in which they try to formalize the trust
values in matrix. They use the matrix multiplication for trust
propagation. The approach is really naive and needs a lot of
enhancements to meet our requirements. Also it is really time
consuming.

\section{Related Works}
Recently, there has been some works on trust propagation, and using
trust information for recommendation. In the following subsection,
we'll review some of the most well known works.
\subsection{Tidal Trust and its application in recommendation} The problem definition
in Jennifer Golbeck's PhD thesis \cite{golbeck_computing} is the
same as the problem we have in mind. She explains an algorithm named
TidalTrust to infer trust and exploits it for recommendation.
TidalTrust is a modified breadth-first search. The source's inferred
trust rating for the sink ($t_{source,sink}$) is a weighted average
of the source's neighbors' ratings of the sink (see the formula).
The source node begins a search for the sink. It polls each of its
neighbors to obtain their rating of the sink. If the neighbor has a
direct rating of the sink, that value is returned. If the neighbor
does not have a direct rating for the sink, the neighbor queries all
of its neighbors for their ratings, computes the weighted average as
shown in formula, and returns the result. Each neighbor repeats this
process, keeping track of the current depth from the source. Once a
path is found from the source to the sink, a depth limit is set.
Since the search is proceeding in a Breadth First Search fashion,
the first path found will be at the minimum depth. The search will
continue to find any other paths at the minimum depth. Once this
search is complete, the trust threshold (max) is established by
taking the maximum of the trust paths leading to the sink. With the
max value established, each node completes the calculations of a
weighted average by taking information from nodes that they have
rated at or above the max threshold. Those values are passed back to
the neighbors who queried for them, until the final result is
computed at the source.

\begin{math}
    t_{i,s} = \frac{\sum_{j\in adj(j) | t_{i,j}\geq max}t_{i,j}t_{j,s}}{\sum_{j\in adj(j) | t_{i,j}\geq max}t_{i,j}}
\end{math}

The "Recommended Rating" is personalized using the trust values for
the people who have rated the film (the raters). The process for
calculating this rating is very similar to the process for
calculating trust ratings in TidalTrust. First, the system searches
for raters that the source knows directly. If there are no direct
connections from the user to any raters, the system moves one step
out to find connections from the user to raters of path length 2.
This process repeats until a path is found. The opinions of all
raters at that depth are considered. Then, using TidalTrust, the
trust value is calculated for each rater at the given depth. Once
every rater has been given an inferred trust value, only the ones
with the highest ratings will be selected; this is done by simply
finding the maximum trust value calculated for each of the raters at
the selected depth, and choosing all of the raters for which that
maximum value was calculated. Finally, once the raters have been
selected, their ratings for the movie (in number of stars) are
averaged. For the set of selected nodes $S$, the recommended rating
$r$ from node $s$ to movie $m$ is the average of the movie ratings
from nodes in S weighted by the trust value $t$ from $s$ to each
node:

\begin{math}
    r_{s,m} = \frac{\sum_{j\in S}t_{s,j}r_{j,m}}{\sum_{j\in S}t_{s,j}}
\end{math}

One thing which should be mentioned about Golbeck's work is ignoring
the profile similarities among users and relying just on trust
network.

\subsection{Works done by Paolo Massa using MoleTrust trust metric }

In Massa's work during his PhD
\cite{massa_moleskiing}\cite{massa_trust}\cite{massa_trust2}, two
matrices are given to the system as input data. The rating matrix
($N\times M$), and trust matrix ($N\times N$). He uses a trust
propagation algorithm (MoleTrust) to infer the indirect trust
values, and the trust matrix will be updated with inclusion of
indirect trust values. Using the rating matrix and Pearson
Correlation, he builds an $N\times N$ matrix for User Preference
Similarity. He uses a snapshot of the dataset and performs the whole
process of calculating the indirect trust values on that snapshot.

MoleTrust predicts the trust score of source user on target user by
walking the social network starting from the source user and by
propagating trust along trust edges. Intuitively the trust score of
a user depends on the trust statements of other users on her (what
other users think of her) weighted by the trust scores of those
users who issued the trust statements. The idea is that the weight
by which the opinion of a user is considered depends on how much
this user is considered trustworthy. Since every trust propagation
starts from a different source user, the predicted trust score of a
certain user A can be different for different source users. In this
sense, the predicted trust score is personalized.

Basically, the MoleTrust trust metric can be modeled in 2 steps.
Step 1 task is to remove cycles in the trust network and hence to
transform it into a directed acyclic graph. Step 2 consists of a
graph walk starting from source node with the goal of computing the
trust score of visited nodes.

Now, for a specified item $i$, they use the following formula to
calculate the predicted rating on it for the user $a$:

\begin{math}
p_{a,i}= \bar{r}_a + \frac{\sum_{u=1}^{k}w_{a,u}(r_{u,i}-
\bar{r}_u)}{\sum_{u=1}^{k}w_{a,u}}
\end{math}

Neighbors can be taken from the User Similarity matrix or from the
Estimated Trust matrix and the weights $w_{a,u}$ are the cells in
the chosen matrix. For example, in the first case, the neighbors of
user $i$ are in the $i$th row of the User Similarity matrix. They
also mention that a combination of these two matrices can be used,
but they don't do that.

The main weakness we see in his approach is the time complexity of
the algorithm he uses. $N$ and $M$ are large values and calculating
the whole trust and similarity values are really time consuming.
Moreover, they do not consider the smallness of the neighborhood of
users which have expressed ratings for that item.

The experimental results show that the improvement of accuracy
comparing to CF algorithms is not very much, but the coverage is
improved by 20\% in their algorithm. The reason for tiny improvement
in accuracy could be ignoring the fact that the number of ratings
for each item is very low in average.

The most interesting part of their approach is the improvement in
coverage for cold start users. He does the whole task of the
recommendation offline (meaning that they just use a snapshot of the
network), and that's the reason he used the smaller basic dataset,
rather than the extended dataset because the time complexity is very
large. It seems that his work is mostly for cold start user, and
improving the coverage for them.

The final point we should mention is the similarity of MoleTrust and
TidalTrust. Both algorithms follow the same idea, but with different
approaches. Actually, our approach will also follow a similar idea
but with a different approach to enhance the efficiency of trust
calculation.

\subsection{Advogato Trust Metric by Levien in UC Berkeley}

The Advogato maximum flow trust metric has been proposed by Levien
and Aiken (\cite{levien_advogato}) in order to discover which users
are trusted by members of an online community and which are not.
Hereby, trust is computed by a centralized community server and
considered relative to a seed of users enjoying supreme trust.
However, the metric is not only applicable to community servers, but
also to arbitrary agents which may compute personalized lists of
trusted peers and not one single global ranking for the whole
community they belong to. In this case, the agent itself constitutes
the singleton trust seed.

The input for Advogato is given by an integer number $n$, which is
supposed to be equal to the number of members to trust, as well as
the trust seed $s$, being a subset of the entire set of users $V$.
The output is a characteristic function that maps each member to a
Boolean value indicating trustworthiness.

Capacities $C_V: V\rightarrow N$ are assigned to every community
member $x\in V$ based upon the shortest-path distance from the seed
to $x$. Hereby, the capacity of the seed itself is given by the
input parameter $n$ mentioned before, whereas the capacity of each
successive distance level is equal to the capacity of the previous
level $l$ divided by the average outdegree of trust edges $e\in E$
extending from $l$. The trust graph obtained hence contains one
single source, which is the set of seed nodes considered one single
"virtual" node, and multiple sinks, i.e., all nodes other than those
defining the seed. Capacities $C_V(x)$ constrain nodes. In order to
apply Ford-Fulkerson maximum integer network flow, the underlying
problem has to be formulated as single-source/single-sink, having
capacities $C_E: E\rightarrow N$ constrain edges instead of nodes.
Hence, following algorithm is applied to the old directed graph $G =
(V, E,C_V)$, resulting in a new graph structure $G' = (V',
E',C_{E'})$.

Eventually, trusted agents $x$ are exactly those peers for which
there is flow from "negative" nodes $x^-$ to the super-sink. An
additional constraint needs to be introduced, requiring flow from
$x^-$ to the super-sink whenever there is flow from $x^-$ to $x^+$.

This approach is interesting. But it needs that we know the whole
structure of the network, which is not feasible in real world
networks. It cannot be run locally, because the transformation of
the network needs the complete knowledge of the network.

\subsection{AppleSeed Trust By Ziegler}
This is the main work Ziegler has done in his PhD thesis
\cite{ziegler_toward}.  In contrast to Advogato, being inspired by
maximum network flow computation, the basic intuition of Appleseed
is motivated by spreading activation models.

The idea is similar to the idea for search in contextual graphs.
Source node s to start the search from is activated through an
injection of energy $e$, which is then propagated to other nodes
along edges according to some set of simple rules: all energy is
fully divided among successor nodes with respect to their normalized
local edge weight, i.e., the higher the weight of an edge $(x, y)\in
E$, the higher the portion of energy that flows along that edge.
Furthermore, supposing average outdegrees greater than one, the
closer node $x$ to the injection source $s$, and the more paths
leading from $s$ to $x$, the higher the amount of energy flowing
into $x$.

In the AppleSeed algorithm, they also have a decay factor $d$.
Hereby, let $in(x)$ denote the energy influx into node $x$.
Parameter $d$ then denotes the portion of energy $d · in(x)$ that
the latter node distributes among successors, while retaining $(1 -
d) · in(x)$ for itself.

One problem we see in this approach is that, they assume the trust
to be additive. Suppose we want to compute the trust from source to
target. There are many weakly trusted paths to target, which
according to their algorithm sums up to high trust value. But, this
is not intuitive.

\subsection{The trust-based recommender proposed by researchers in ETHZ}
This research \cite{walter_model} is the most recent work in this
field. In this work, they present a model of a trust-based
recommendation system on a social network. The idea of the model is
that agents use their social network to reach information and their
trust relationships to filter it. They investigate how the dynamics
of trust among agents affect the performance of the system by
comparing it to a frequency based recommendation system.

Their model consists of Agents, objects, and profiles. When facing
the purchase of an item, agents query their neighborhood for
recommendations on the item to purchase. Neighbors in turn pass on
a query to their neighbors in case that they cannot provide a reply
themselves. In this way, the network replies to a query of an
individual by offering a set of recommendations. One way to deal
with these recommendations would be to choose the most frequently
recommended item. However, because of the heterogeneity of
preferences of agents, this may not be the most efficient strategy
in terms of utility. Thus, they explore means to incorporate
knowledge of trustworthiness of recommendations into the system.

They use the discrete values of 1 and -1 for agent's ratings over
items. They also use a na\"{\i}ve approach to propagate the trust in
network to infer indirect trusts. They just multiply the trust
values along the path between the source and target agent. We can
identify two problems with this approach. First of all, in a path
between the source and target agent, the edge closer to the source
should have more impact on the indirect trust value. Second, what if
there are multiple paths between source and target?

For deciding what items to recommend they find the probability for
recommending each item among the set of selected items. Then they
recommend each item with the probability associated with it. It
seems a little weird to us to have different recommendations for an
agent in two consecutive queries for recommendations. Also we
believe user acceptance of this approach is not very good since they
feel inconsistency among recommendations. What they do is that they
find a probability for each item to be recommended. Then they sample
a set of items to be recommended from the whole items according to
the probabilities associated with each item. Maybe explaining the
probabilities of each item in a user friendly manner would be more
appropriate.

\section{Our Proposed Method}

As discussed in previous section, there are some issues with each
approach in literature which should be dealt in a smart way. We
briefly review the problems with each approach in the following.

\begin{itemize}
  \item MoleTrust Used by Massa
    \begin{itemize}
        \item This approach works just on a snapshot of the network.
        So this approach can not catch the network evolution and
        updated in trust values.
        \item The time complexity of this approach is very high. For
        each query, we have to multiply big matrices.
    \end{itemize}
  \item TidalTrust by Golbeck
    \begin{itemize}
        \item This approach also has the problem of time complexity.
        For each query, we should traverse a huge part of the network
        to find the appropriate node having the rating. The reason
        is that for each user and each item we have to look among
        the whole users at a certain depth of the network from
        user's point of view.
        \item In this approach, when we reach a node having the
        rating, we just consider the nodes at this depth. This is a
        very strict constraint, which may lead to have just one node
        having the rating (at that depth).
    \end{itemize}
  \item Advogato by Lenien
    \begin{itemize}
        \item The parameter n is an extra input which is hard to
        tune.
        \item This approach just recommend which users to trust.
        There is no trust value associated with users.
        \item We have to have a complete knowledge of the network to
        be able to assign the capacities for each query.
    \end{itemize}
  \item AppleSeed by Ziegle
    \begin{itemize}
        \item This approach assumes that trust is additive.This is a
        incorrect assumption. Suppose we want to compute the trust from source to
        target. There are many weakly trusted paths to target, which
        according to their algorithm sums up to high trust value. But, this
        is not intuitive.
    \end{itemize}
  \item The model proposed by researchers in ETHZ
    \begin{itemize}
        \item They do not consider a damping factor for trust along
        the path. This is also a problem with Massa's approach and
        Golbeck's Tidal Trust.
        \item The way they output the result based on the
        probability of each item to be recommended is weird. User can not accept different result
        for the same query.
    \end{itemize}
\end{itemize}

To deal with these issues and having an approach which has as few
issues as possible, we propose a distributed approach which will be
explained in the following paragraphs.

Generally, we use the following approach in our first project. The
approach is in two steps:
\begin{enumerate}
  \item We find the neighborhood of trusted users who have ratings for the item.
  This neighborhood contains users $U_n=\{u_1, ... u_k\}$. and for each user $u_i$, we have the inferred trust value $t_i$.
  \item Now, we should aggregate the ratings from different users in neighborhood to find a recommended rating for the item
\end{enumerate}

As discussed in related work section, the approaches used by Golbeck
and Massa follow similar ideas. They just differ in minor details.
What we have in mind is very similar to their idea. The essential
consideration in this approach is locally feasibility of this
approach. Let's first explain the approach.

Suppose we have a user $X$ and another user $Y$. We want to compute
the trust from $X$ to $Y$. Then we define $N_{(X,Y)}=\{i\in direct\
neighbours\ of\ X |  trust_{(X,i)}>0 and\ i\ has\ trust\ values\
for\ Y\}$. The constraint of trusts in neighbors to be positive is
for distrust values. In our approach we try to also take distrust
into account (Although our data set does not contain distrust
values). Distrust propagation has received very few attention in
literature, and none of the approaches described in related works
section consider distrust. For distrust propagation we just consider
distrust to Y from trusted neighbors of X; because distrust values
acquired by distrusted neighbors are meaningless.

Now, the trust value from $X$ to $Y$ would be:
\begin{equation}
trust_{(X,Y)}= \frac{\sum_{i\in N_{(X,Y)}}trust_{(X,i)}\lambda
trust_{(i,Y)}}{\sum_{i\in N_{(X,Y)}}trust_{(X,i)}}
\end{equation}

The damping factor $\lambda$ penalizes the long paths to $Y$. At the
beginning, only the trust values for direct links are set, and not
all direct neighbors have trust values for $Y$.

So, what we do is, before asking the neighbors about their trust
values on $Y$ we do an iterative procedure to augment the network so
that each user also has the indirect trust values. This procedure
can be done periodically to maintain the network up to date. In each
iterations, each user asks its direct neighbors for their most up to
date neighborhood (both direct and indirect). Aggregating these
trust values, the user will update its neighborhood to catch the
most up to date changes in trust network. The trust from indirect
neighbors will be gradually propagated in iterations.

To accomplish the above mentioned procedure, each node should keep
track of its direct and indirect neighbors. Also it should store the
pointer to node to which it has trust expressions. This needs some
resource, which can be handled by applying threshold on trust of
trusted users. The threshold could be a user defined threshold on
the trust to neighbors. Since the information required for each node
is around 3-4 bytes, the total extra resource required would be just
a couple of megabytes, which is worth the advantage of being fast in
responding.

After finding the trust values (which make take some iterations for
convergence of the trust values), we have the whole indirect trust
values. When a new direct trust value is inserted in to the network,
 the new trust value will be propagated through the network when
 updating the network in iterations. So, one important advantage of this algorithm is being adaptable
to network evolution and new links and ratings in the network.

As you can see, the idea is similar to idea used in TidalTrust and
MoleTrust, but the approach to implement is more efficient so that
the procedure can even be implemented in parallel. This idea is also
very useful in peer to peer network. The efficiency we claim is on
time efficiency. As stated by Massa, it took 7 days for him to
perform the experiments, which is quite a long time for a real
application. We believe by sacrificing some space, we could getter
much better time efficiency.

The most important motivation for using this approach is its
application in distributed networks. In distributed networks, there
is no central database to make the recommendation, and each user has
local access to its neighbors only. So using an iterative approach
to store the indirect neighbors makes a lot of sense in distributed
network for efficiency. In other word, our approach is similar to
Massa's approach in the sense that it tries to compute the trust
between all pairs. But unlike Massa's approach, we do it in
parallel, and store the trust values in distributed resources,
rather than a central matrix.

\begin{center}
\begin{figure}
  \includegraphics[width=7cm]{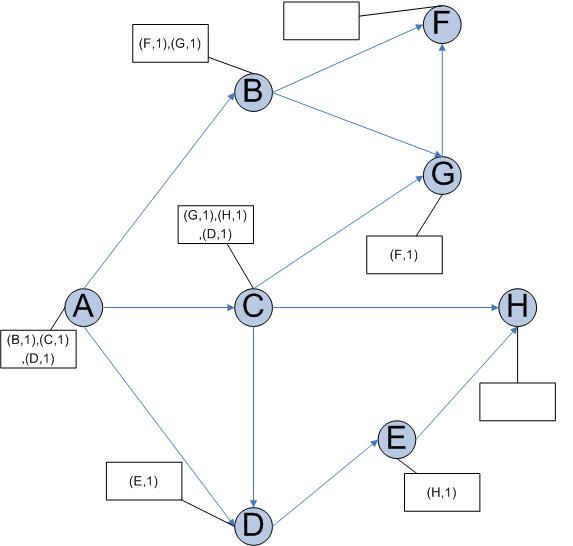}\\
  \caption{The trust network before running any iteration}\label{proposed1}
\end{figure}
\end{center}

\begin{center}
\begin{figure}
  \includegraphics[width=7cm]{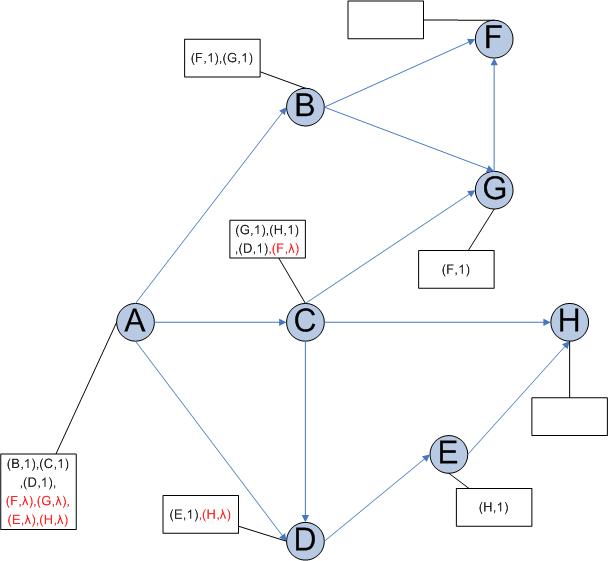}\\
  \caption{The trust network after first iteration}\label{proposed2}
\end{figure}
\end{center}

\begin{center}
\begin{figure}
  \includegraphics[width=7cm]{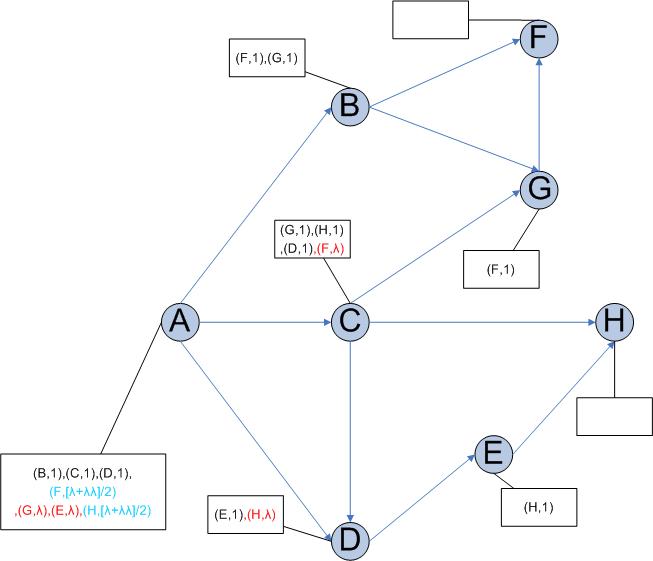}\\
  \caption{The trust network after second iteration. This is the final iteration}\label{proposed3}
\end{figure}
\end{center}

To clarify our proposed method, let's discuss an example. Suppose we
have a trust network as shown in figure \ref{proposed1}. The boxes
in the figure show the neighborhood for each user. Each pair in the
box shows a neighbor and the trust value for that neighbor. At the
beginning, the neighborhood is just the direct neighbors. Running
the first iteration, the neighborhoods would be updated as shown in
figure \ref{proposed2}. The pairs shown in red are the new trust
values inserted in the network. Now, if we run the second iteration,
we'll get the network as shown in figure \ref{proposed3}. The pairs
in blue are trust values updated in this iteration. This is the
final iteration, and running any more iteration will not change the
network.

Now, having the enhanced network, to recommend a rating for an item
$a$, we use the following procedure:

$N_{(X,a)}=\{i\in memmbers\ of\ neighbourhood\\ who\ have\ ratings\
for\ item\ a\}$
\begin{multline}
RecommendedRating_{X,a}=\\ \frac{\sum_{Y\in
N_{X,a}}trust_{(X,Y)}\times r_{Y,a}}{\sum_{Y\in
N_{X,a}}trust_{(X,Y)}}
\end{multline}
We can also define a confidence value, which shows how confidence we
are for our rating as follows:
\begin{multline}
Confidence_{X,a}= \frac{Average_{y\in
N_{X,a}}\{trust_{(X,Y)}\}}{Variance_{y\in N_{X,a}}\{r_{Y,a}\}}
\end{multline}

\section{Evaluation}
There is a general approach in recommendation which can be used in
our approach. This approach is named "leave-one-out". In this
approach, we omit one of the ratings from a user profile, and ask
the system to predict the rating. Precision and recall can be used
for accuracy of prediction.

Also the same approach can be used for trust propagation. We can
just remove one trust expression (an edge in the graph), and ask the
system to predict the trust value based on the rest of the network.
The definition of precision is straightforward in this context.

For recall, notice that some preferences or trust values cannot be
predicted. So the recall value is simply the fraction of trust
values which can be predicted.

To have some works to compare our work with those, we will also
implement the works done by Golbeck and Massa.

\section{Data Set Specification}
The dataset was crawled from Epinions website by Paolo
Massa\footnote{http://www.trustlet.org/wiki/Downloaded\_Epinions\_dataset}
in his PhD thesis. The dataset contains
\begin{itemize}
  \item 49,290 users who rated a total of
  \item 139,738 different items at
least once, expressing
  \item 664,824 ratings.
  \item 487,181 issued trust
statements.
\end{itemize}
According to the above statistics, we can infer the following
results:
\begin{itemize}
  \item Each user has 10 neighbors in average.
  \item Each user has expressed 15 ratings in average.
  \item Each item has around 5 ratings in average.
\end{itemize}
The first two results are reasonable. But, the third one is really
surprising. It means that, on average, the neighborhood for each
user for a specified item is at most 5 (ignoring the trust threshold
for neighbors). This will cause problems for our approach. Since we
are first building the neighborhood and then using the neighborhood
to predict the rating, small neighborhoods cannot help us. They just
increase the time complexity of our neighborhood building procedure.
That's the reason we proposed an iterative approach which leads to
finding the indirect neighbors very fast.

\section{Experiments with TidalTrust}
We implemented the TidalTrust algorithm using Java 1.6. The DBMS we
used is MySQL 4.1.16.
\\
While running the algorithm, we face some problems. In Golbeck's
algorithm we search the graph in a breadth first manner to find a
user having the rating for the specified item. This user is at a
depth d. Now, if the depth is 1, the algorithm finds the result in
about 150ms. But, as soon as for a recommendation query we need to
go further in depth two, the running time would be around 5 seconds.
This gets exponentially worse when $d=3$. In this case it takes 15
minutes in average to compute the results, which is not acceptable.
Obviously the problem would be worse for bigger values of $d$.
Apparently, one reason for this problem is Java, and how Java
handles connection to database. (We even tried changing the database
and working on SQLServer, but the problem got worse there).
\\
But still, for the case $d=3$ , the algorithm needs around 50000
queries to database, which seems unacceptable for a single
recommendation query for a pair of user and item.
 So, we designed a Cache, which cached each query to database.
 In this case after a while there was no need to access database;
 all we needed was already in the Cache. Notice that we could just turn cache
  off to directly communicate with database in each recommendation query.
  But for a experiment with more than 600000 pairs of <user,item>, it would take like one year.
  Even with using the cache, it took 72 hours of continuous running of the application with CPU usage of 50\%,
  and memory usage of 500MB.
\\
  In the following subsections, we first discuss some basic results.
  Then we evaluate the result with approaches presented in Golbeck
  and Massa's works.

  \subsection{General results}

\begin{center}
\begin{figure}
  \includegraphics[width=6cm]{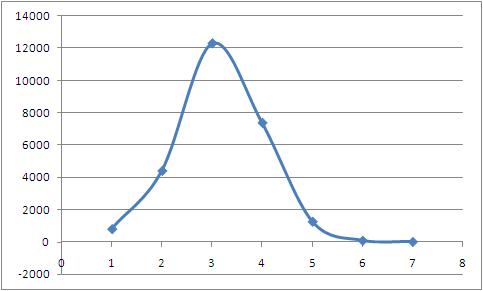}\\
  \caption{The distribution of the max-depth among users.}\label{fig-dist-max-depth}
\end{figure}
\end{center}

Our data set just contains binary trust values.
   Since the TidalTrust does not have any damping factor,
   the recommended rating would be the average of ratings
   from users at a specified depth. So, the TidalTrust
   algorithm does not work properly in this data set.
   Also, the number of users having the rating at the specified depth
   is usually a percentage of the whole users having the rating, which leads to losing some information.
   Lacking the damping feature is a big issue in Golbeck's work
   which can be resolved by applying damping factor.
   \\
   Table 1 shows the distribution of the depth at which we find a rating
   for that item. Depth=-1 means that we could not find any other ratings
   for that item. This is one potential drawback to creating recommendations based
   solely on relationships in the social network is that a recommendation
   cannot be calculated when there are no paths from the source to any
   people who have rated a movie. This case is rare, though, because as long as
   just one path can be found, a recommendation can be made. In the FilmTrust
   network, when the user has made at least one social connection, a recommendation
   can be made for 95\% of the user-movie pairs. But, as you can see in table 1, 24.35\% of <user,rating>
    pairs were unique, and the algorithm was not able to recommend a rating for that pair, which is much
    worse than the case in FilmTrust network. This is so called the
    coverage of the ratings which can be predicted by the system (75.65\%).
    We'll discuss the coverage in detail later in this paper.\\
{\small
\begin{table}[h]
\begin{center}
\label{tab:dist-depth}
\begin{tabular}{|c|c|}
\hline
Depth& Ratings\\
\hline \hline
-1&  157392 \\
\hline \hline 1&  175124
 \\
\hline \hline 2& 208088
  \\
\hline \hline 3&  86190
 \\
\hline \hline 4&  17463
 \\
\hline \hline 5&   1960
\\
\hline \hline 6&   130
\\
\hline \hline 7& 15
  \\
\hline
\end{tabular}
\caption{Distribution of the depth the item is found for a simple
query}
\end{center}
\end{table}}
Also, this table can be some used to approximate the diameter of the
network. According to this table most of the paths in the network
are at most of length 5, which leads to a diameter of five. This
diameter is less than the general diameter for social networks (6).
This means that this network is a dense network. To clarify the
implication of the diameter, we define a metric called max-depth.
Max-depth is the maximum depth a user needs to investigate to find a
rating for on the item it rates. Figure \ref{fig-dist-max-depth}
shows the distribution of max-depth among users. This figure also
shows that most users find the rating for an item in depth less than
or equal to six.

\begin{center}
\begin{figure}
  \includegraphics[width=7cm]{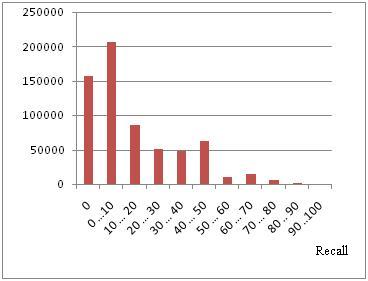}\\
  \caption{The detailed distribution of the rating-recall. The average recall is 19.36\%}\label{fig-dist-recall}
\end{figure}
\end{center}

As mentioned earlier, the nature of the TidalTrust will lead to
losing some of the ratings for an item. Because all the ratings for
an item are not at the same depth, and this leads to ignoring some
of them. For each recommendation query, we define a rating-recall
metric which is equal to the percentage of ratings for that item
considered for the recommendation. Table 2 and figure
 \ref{fig-dist-recall} show the the
distribution of the recall. According to the diagram, the average
percentage of ratings considered to recommend a rating is 19.36\%
which is a pretty low percentage and shows information loss.

{\small
\begin{table}[h]
\begin{center}
\label{tab:dist-recall}
\begin{tabular}{|c|c|}
\hline
Recall Percentage& count\\
\hline \hline 0&   157392
\\
\hline \hline 0 .. 10& 205787
  \\
\hline \hline 10 .. 20&  86409
 \\
\hline \hline 20 .. 30& 51308
  \\
\hline \hline 30 .. 40&  48459
 \\
\hline \hline 40 .. 50&   63108
\\
\hline \hline 50 .. 60&  9935
 \\
\hline \hline 60 .. 70&  14047
 \\
\hline \hline 70 .. 80&  5981
 \\
\hline \hline 80 .. 90&   942
\\
\hline \hline 90 .. 100&  22
 \\

\hline
\end{tabular}
\caption{The distribution of the rating-recall. The average recall
is 19.36\%}
\end{center}
\end{table}}

\subsection{Experimental Results for TidalTrust}
As mentioned in Evaluation section, the main approach for evaluation
the result is the "leave-one-out" method. In this case, for each
rating expressed by a user, we'll have an absolute error which is
the difference between the actual rating and the recommended rating.
Sine the ratings are in range [1..5], the absolute error value range
from zero to 4.
\begin{center}
\begin{figure}
  \includegraphics[width=7cm]{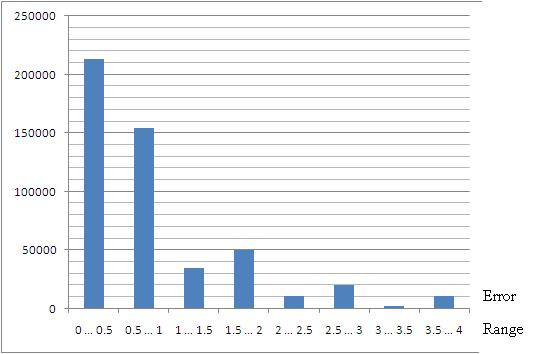}\\
  \caption{The distribution of errors in TidalTrust Experiment.}\label{fig-dist-error}
\end{figure}
\end{center}
Figure \ref{fig-dist-error} shows the distribution of errors in
TidalTrust Experiment. As shown in the diagram, most of errors are
in range [0..1] which looks like a promising result. But, as
discussed in Golbeck's thesis \cite{golbeck_computing}, this is just
because of the nature of the social networks, in which people tend
to rate items close to the average rating for that item.

Golbeck mentions that the point of the recommended rating is more to
provide useful information to people who disagree with the average.
In those cases, the personalized rating should give the user a
better recommendation, because we expect the people they trust will
have tastes similar to their own. The difference between the user's
actual rating and the average rating is called $\Delta a$. Users who
disagree with average have large values of $\Delta a$ for their
rating. She also defines $\Delta r$ as the difference between the
actual rating and the recommended rating. As the base method, she
uses Automatic Collaborative Filtering (ACF). So, she defines
$\Delta cf$ as the difference between a user's actual rating of a
film and the ACF calculated rating.

\begin{center}
\begin{figure}
  \includegraphics[width=7cm]{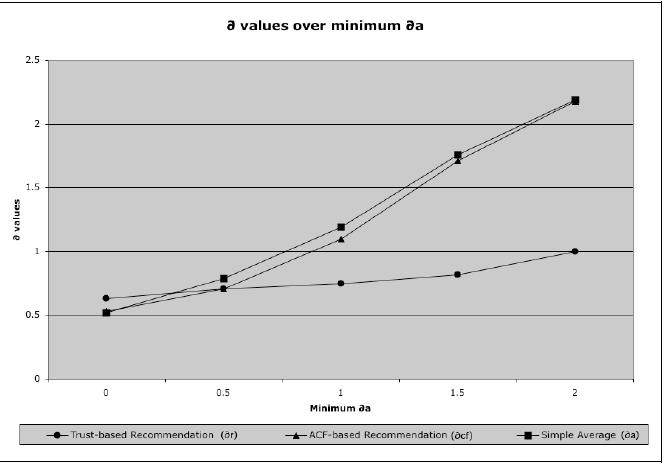}\\
  \caption{The increase in $\Delta$ as the minimum $\Delta a$ is
increased. Notice that the ACF-based recommendation ($\Delta cf$)
closely follows the average ($\Delta a$). The more accurate
Trust-based recommendation ($\Delta r$) significantly outperforms
both other methods.}\label{fig-golbeck}
\end{figure}
\end{center}

Defining a threshold on the $\Delta a$ of ratings being considered
in evaluation, we'll get different sets of users to be considered in
our evaluation. Golbeck compared three recommendation methods
(TidalTrsut, Simple Average, and ACF) considering these threshold as
shown in figure \ref{fig-golbeck} (taken from her thesis),
TidalTrust works much better than the other two methods when the
threshold (minimum $\Delta a$ gets larger). These results are for
the data set FilmTrust.

\begin{center}
\begin{figure}
  \includegraphics[width=7cm]{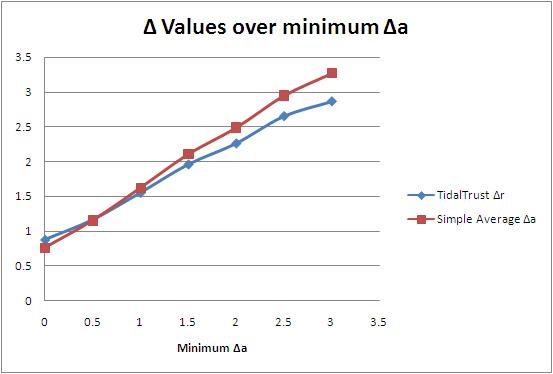}\\
  \caption{The increase in $\Delta$ as the minimum $\Delta a$ is
increased. TidalTrust closely follows the Simple Average algorithm
in Epinions  data set}\label{fig-delta-ours}
\end{figure}
\end{center}

We applied the same evaluation metric for TidalTrsut on our
experiment. Figure \ref{fig-delta-ours} show the results of the
experiments of TidalTrust on Epinions. Although, as mentioned
before, TidalTrust on Epinions works more or less like averaging.
But since it looses some of the ratings (which Golbeck claimed these
items are far away from the user and should not be considering
-without taking the trust value of the path into account-), we
compared the two algorithms. As you can see in the figure, the
difference between TidalTrust is very low on all thresholds. So
TidalTrust does not work well on Epinions.
One point is that TidalTrust is an algorithm introduce for
continuous trust values, not for binary trust values. But, even the
algorithm introduced in \cite{golbeck_computing} for binary trust
values will work in the same way. The problem with these approaches
is not considering the damping of trust. They do not consider the
length of the path from one user to another user.

Paolo Massa, who has his own trust based recommendation system
\cite{massa_trust}, uses a somehow different approach for evaluation
the recommendation. To decrease the influence of users with a lot of
ratings on Mean Absolute Error (MAE), he uses Mean Absolute User
Error (MAEU) metric. The idea is straightforward: we first compute
the Mean Absolute Error for every single user independently and then
we average all the Mean Absolute Errors related to every single
user. In this way, every user has the same weight in the Mean
Absolute User Error computation. This is really important since
Epinions dataset contains a large share of cold start users
\cite{massa_trust}.

Another important measure that is often not reported and studied in
evaluation of RSs is coverage. Coverage simply refers to the
fraction of ratings for which, after being hidden, the RS algorithm
is able to produce a predicted rating. While the percentage of
predictable ratings (\emph{ratings coverage}) is an important
measure, it suffers the same problem we highlighted earlier for Mean
Absolute Error, it weighs heavy raters more. Following the same
argument as before, Massa introduced also the \emph{users coverage},
defined as the portion of users for which the RS is able to predict
at least one rating \cite{massa_trust}.

A possibility given by a very large data set of ratings is to study
performances of different RS techniques on different portions of the
input data (called "views"). It is possible for example to compute
MAE or Users coverage only on ratings given by users or items which
satisfy a certain condition. The views Massa reported results about
in his paper are the following: cold start users, who provided from
1 to 4 ratings; heavy raters, who provided more than 10 ratings;
opinionated users, who provided more than 4 ratings and whose
standard deviation is greater than 1.5; niche items, which received
less than 5 ratings; controversial items, which received ratings
whose standard deviation is greater than 1.5. They introduced these
views because they are better able to capture the relative merits of
the different algorithms in different situations \cite{massa_trust}.

Table 3 shows the evaluation measures for different types of users
introduced above.

{\small
\begin{table*}[h]
\begin{center}
\label{tab:dist-depth}
\begin{tabular}{|c|c|c|c|c|}
\hline
Views& Rating Cov. & User Cov. & MAE & MAUE\\
\hline \hline All
  &0.756
 & 0.652
& 0.874 & 0.58
  \\
\hline \hline Cold Start
 & 0.496
& 0.478 & 0.905 &  0.44
 \\
\hline \hline Heavy Raters
 & 0.787
& 0.844 & 0.871 &  0.74
 \\
\hline \hline Opti. Rater
 & 0.753
& 0.745 & 1.137 &  0.86
 \\
\hline \hline Niche Items
 & 0.461
& 0.624 &0.845
 &  0.54
 \\
\hline \hline Cont. Items
 &0.853
 & 0.795
& 1.714 &  1.35
 \\
\hline
\end{tabular}
\caption{User Coverage, Rating Coverage, MAE, and MAUE on different
views of the users and items}
\end{center}
\end{table*}}

The results of TidalTrust and MoleTrust should be almost the same.
Theoretically, the idea being used in both algorithms is the same.
But they use different approaches. In TidalTrust, the aim is to find
the trust from the source user to the target user. To compute this
trust value, TidalTrust uses a recursive approach: Starting from the
source user ($u$), the trust to target user ($v$) is the weighted
average of the trust of $u$'s neighbors to $v$. On the other hand,
in MoleTrust, we try to find trust values into all nodes in the
network. The trust into the source ($u$) is 1. Then, we traverse the
graph in a breadth first manner (level by level). Now, the trust
into each node $w$ (which lies in level $i$) is the weighted average
of trust into its inlinks in level $i-1$. Comparing the
recommendation systems based on these two approaches, the method is
the same. But, there are some minor differences in how to select the
neighbors. TidalTrust just considers the neighbors at a certain
depth which leads to information loss in most cases. On the other
hand, MoleTrust revises the network and considers the graph as some
level of nodes starting from the source user. But generally, these
two approaches work the same.

But, compare to what shown in Massa's paper \cite{massa_trust}, the
coverage for users and items are mostly lower in TidalTrust
comparing to Massa's work, except for controversial items. Actually
the coverage for all user views in TidalTrust are close (because of
the averaging nature of it), and so it doesn't differentiate among
different views.

One interesting point comparing the results of TidalTrust and
Massa's work is better errors in TidalTrust. The reason is maybe
because most people tend to rate items close to the average.

An important issue existing in both TidalTrust and MoleTrust is the
number of queries to the database for every single recommendation
query. Table 4 shows the average queries required for each
recommendation according to the depth $d$ at which the item is
found. As you can see in the table, the average number of queries to
database is too much. This huge number of queries slow down the
process of recommendation. Using our proposed method will help
decrease the amount of queries.

{\small
\begin{table}[h]
\begin{center}
\label{tab:queries}
\begin{tabular}{|c|c|}
\hline
Depth & Average Queries\\
\hline \hline
1 &  95.0837 \\
\hline
    2 & 8169.4467 \\
     \hline
      3 &  23148.7689 \\
      \hline
       4 & 43042.2457 \\
        \hline
5 &   58528.1933 \\
\hline
 6 &   76093.8083 \\
 \hline
    7 &   85910.6452 \\
    \hline
         8 &  99066.7143 \\
         \hline
   9 &  106309.0000 \\
   \hline
\end{tabular}
\caption{Average queries required for each recommendation according
to the depth $d$ at which the item is found.}
\end{center}
\end{table}}

\subsection{Experimental Results of our proposed method}

To simulate the parallelism in our proposed method, we implemented
User which can perform all tasks, and has access to only its
neighbors. According to our method, we have to assign values to
parameters. Trust threshold could be a user defined parameter. In
our experiment we set it to 0.7. Damping factor $\lambda$ is also
set to 0.8.

Table 5 shows the comparison of different evaluation measures in
TidalTrust and our proposed method. As we expected, the error is
much less in our proposed method, since we use the information
gathered from just trusted neighbors. But, the coverage is lower.
Because we ignore many low trusted neighbors, the coverage will
decrease. But this is the cost we pay to get better accuracy.

The rating-recall is almost the same. Because both approaches loose
some information. But the loss in our approach is meaningful, since
we get rid od non-trusted information.

{\small
\begin{table}[h]
\begin{center}
\label{tab:queries}
\begin{tabular}{|c|c|c|}
\hline
Metric&TidalTrust & Our Algorithm\\
\hline \hline
Error &  0.92 &  0.54 \\
\hline
Coverage &  75.6\% &  31.2\% \\
\hline
Rating-Recall &  19.36\% &  11.43\% \\
\hline

\end{tabular}
\caption{Comparison of Evaluation measures in TidalTrust and our
proposed algorithm}
\end{center}
\end{table}}

Finally, the average size of neighborhood 1663. So, on average we
need to store the information for 1663 trusted neighbors, which is
not a lot for users.
\section{Conclusions}
In this research project, we first reviewed existing methods for
trust based recommendations. We then proposed a new method which is
iterative and saves a lot of time sacrificing some resource. Our
approach can be easily paralleled and used in distributed networks
in which users have just local access to information.
\bibliographystyle{acl}
\bibliography{Ref}

\end{document}